\documentclass[twocolumn,showpacs,aps,superscriptaddress]{revtex4-1}
\usepackage{amssymb}
\usepackage{amsmath}
\usepackage{graphicx}
\usepackage{epsfig}
\usepackage{xcolor}

\begin{document}

\title{Dynamic transition from insulating state to $\eta $-pairing state in
a composite non-Hermitian system}
\affiliation{School of Science, Chongqing University of Posts and
Telecommunications, Chongqing, 400065, China }
\affiliation{School of Physics, Nankai University, Tianjin 300071, China}
\affiliation{Institute for Advanced Sciences, Chongqing University of Posts and Telecommunications, Chongqing, 400065, China}
\author{X. M. Yang}
\affiliation{School of Science, Chongqing University of Posts and
Telecommunications, Chongqing, 400065, China }
\affiliation{Institute for
Advanced Sciences, Chongqing University of Posts and Telecommunications,
Chongqing, 400065, China}
\author{Z. Song}
\email{songtc@nankai.edu.cn}
\affiliation{School of Physics, Nankai University, Tianjin 300071, China}

\begin{abstract}
The dynamics of Hermitian many-body quantum systems has long
been a challenging subject due to the complexity induced by the
particle-particle interactions. In contrast, this difficulty may be avoided
in a well-designed non-Hermitian system. The exceptional point (EP) in a
non-Hermitian system admits a peculiar dynamics: the final state being a
particular eigenstate, coalescing state. In this work, we study the dynamic
transition from a trivial insulating state to an $\eta $-pairing state in a
composite non-Hermitian Hubbard system. The system is consisted
of two subsystems A and B, which is connected by unidirectional hoppings. We
show that the dynamic transition from an insulating state to an $\eta $%
-pairing state occurs by the probability flow from A to B: the initial state
is prepared as an insulating state of A, while B is left empty. The final
state is $\eta $-pairing state in B but empty in A. Analytical analyses and
numerical simulations show that the speed of relaxation of off-diagonal
long-range order (ODLRO) pair state depends on the order of the EP, which is
determined by the number of pairs and the fidelity of the scheme is immune
to the irregularity of the lattice.
\end{abstract}

\maketitle

\section{Introduction}

Experimental advances in atomic physics, quantum optics, and nanoscience
have made it possible to realize artificial systems. It is fascinating that
some of them are described by Hubbard model \cite{Hubbard} to a
high degree of accuracy \cite{Jochim, Greiner}. Then one can experimentally
realize and simulate the physics of the model. The Hubbard model is a simple
lattice model with particle interactions and has been intensely investigated
in various contexts ranging from quantum phase transition \cite%
{Matthew,Antoine} to high temperature superconductivity \cite%
{Keimer,Patrick,CNY}. Direct simulations of such a simple model is not only
helpful to solve important problems in condensed matter physics, but also to
the engineering design of quantum devices. Importantly, the availability of
experimental controllable Hubbard systems provides an unprecedented
opportunity to explore the nonequilibrium dynamics in interacting many-body
systems.

Very recently, it has been demonstrated that nonequilibrium many-body
dynamics provides an alternative way to access a new exotic quantum state
with energy far from the ground state \cite%
{Choi,Else,Khemani,Lindner,Kaneko,Tindall,YXMPRA,ZXZPRB2}. It makes it
possible to design interacting many-body systems that can be used to prepare some desirable many-body quantum states in principle.
Unlike traditional protocols based on cooling down mechanism, %
quench dynamics has a wide range of potential applications, since it
provides many ways to take a system out of equilibrium, such as applying a
driving field or pumping energy or particles in the system through external
reservoirs \cite{QD1, QD2,QD3}. In recent work Ref. \cite%
{YXMPRB}, a scheme has been proposed to realize quantum mold casting, i.e.,
engineering a target quantum state on demand by the time evolution of a
trivial initial state. The underlying mechanism is pumping fermions from a
trivial subsystem to the one with topological quantum phase. In this work,
we extended this approach to interacting many-body systems.

\begin{figure*}[tbp]
\includegraphics[ bb=42 502 586 790, width=0.495\textwidth, clip]{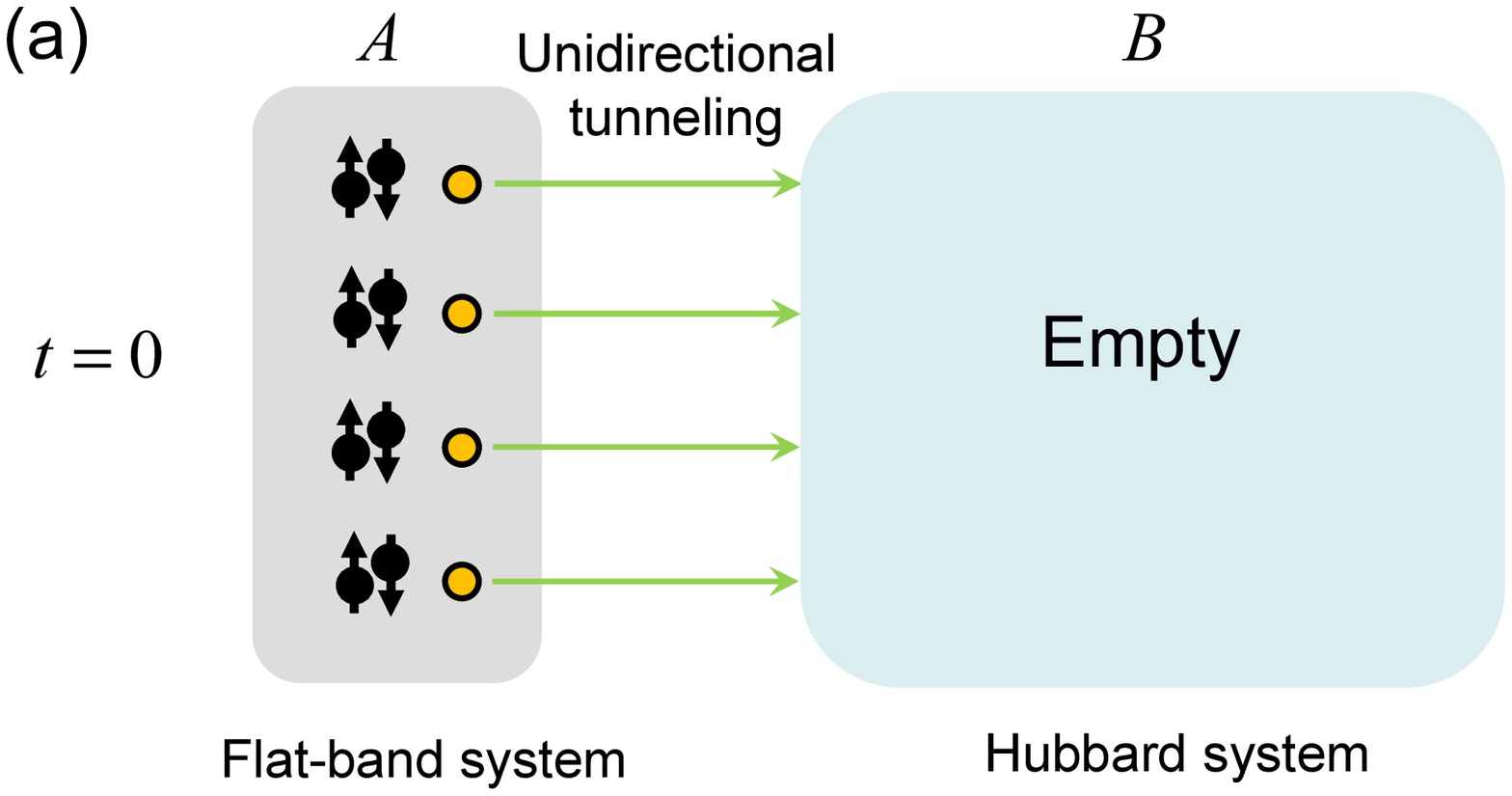} %
\includegraphics[ bb=42 502 586 790, width=0.495\textwidth, clip]{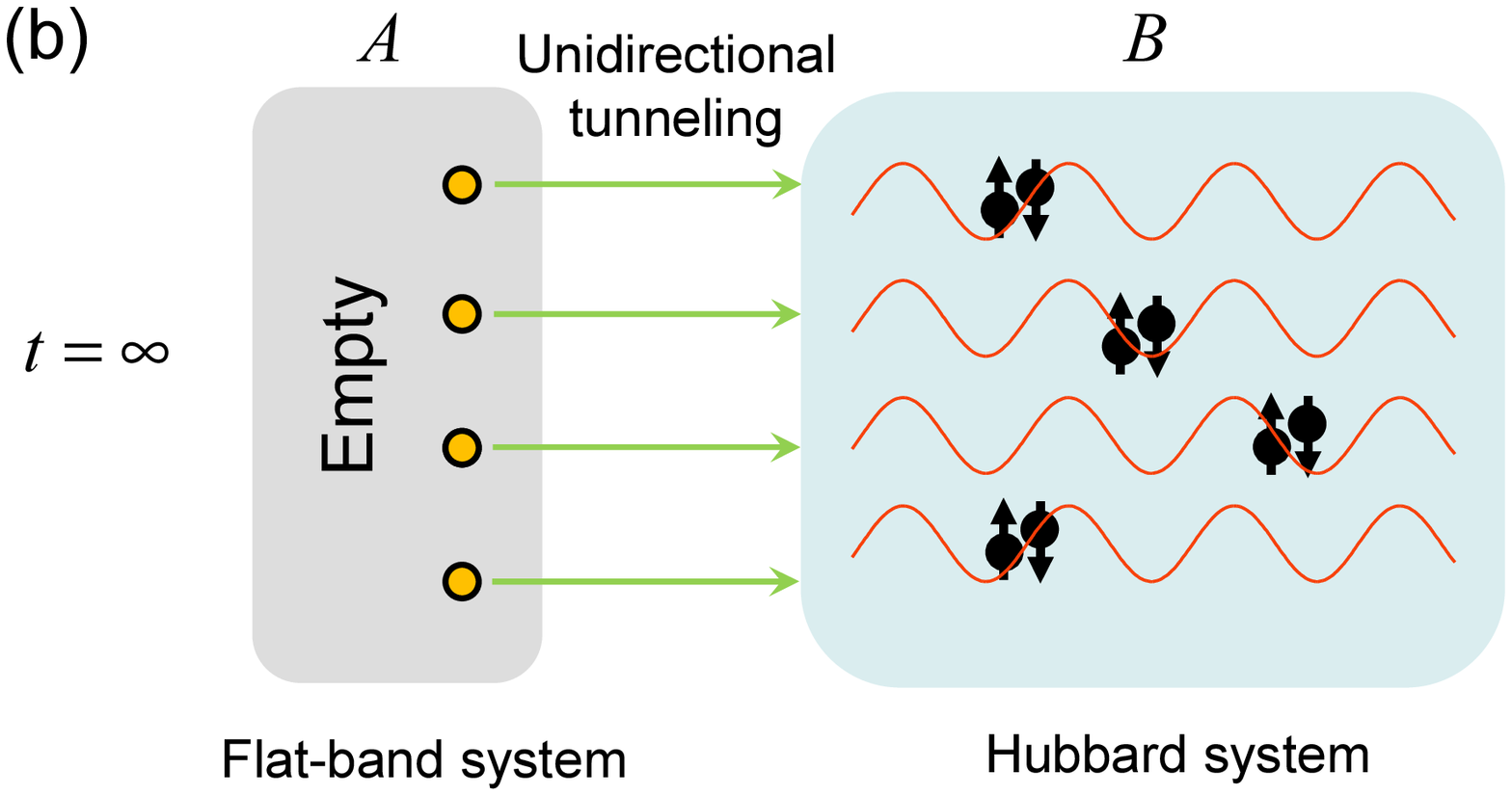}
\caption{The system is consisted of two subsystems A and B, which is
connected by unidirectional hoppings. A is a interaction-free system with
trivial flat-band, while B is a Hubbard system, which supports $\protect\eta
$-pairing eigenstates. At $t=0$, A is fully filled as trivial insulating
state and B is empty. When the unidirectional hoppings switch on, the
probability flow from A to B. After long time, subsystem A becomes empty and
subsystem B is in an $\protect\eta $-pairing state, a condensation of bound
pairs.}
\label{fig1}
\end{figure*}

In general, the time dynamics of Hermitian many-body quantum systems has
long been an elusive subject, due to the complexity induced by the
particle-particle interactions. The main obstacle is that the
evolved state is not easily predictable in most cases. Nevertheless, this
difficulty may be avoided in a well-designed non-Hermitian system, since the
EP in a non-Hermitian system admits a peculiar dynamics: the final state
being a particular eigenstate, coalescing state \cite%
{Berry2004,Heiss2012,Miri2019,Zhang2020,JL}. The key point is the
exceptional dynamics, which allows particles pumping from the source
subsystem to the central subsystem, realizing the dynamical preparation of
many-body quantum states. In present work, we study the dynamic transition
from a trivial insulating state to an $\eta $-pairing state in a composite non-Hermitian Hubbard system. The system is consisted of two
subsystems A and B, which is connected by unidirectional hoppings.\ {Based
on the performance of the system at EP, a scheme that produces a
nonequilibrium steady superconducting-like state is proposed. Specifically,
for an initial state with fully filled in A, but empty in B, unidirectional
hoppings can drive it to the resonant coalescing state that favors
superconductivity manifested by the ODLRO. Such a dynamical scheme can be
realized no matter how the shapes of the two sublattices of the composite
system. Therefore, our finding is distinct from the previous investigations
\cite{Kaneko,Tindall,ZXZPRB2}, and provides a quantum casting mechanism for
generating superconductivity through nonequilibrium dynamics. On the other
hand, the remarkable observation from our work can trigger further studies
of both fundamental aspects and potential applications of composite
non-Hermitian many-body systems.}

The rest of this paper is organized as followed. In Section \ref{model}, we
present the model and its properties relating to $\eta $ operators, or $\eta
$-symmetry. Section \ref{Doublon effective Hamiltonian} is devoted to
doublon effective Hamiltonian which captures the physics in a fixed energy
shell. In Section \ref{Jordan form with high-order EP}, we present the
Jordan form with high-order EP based on the effective Hamiltonian. In
Section \ref{Dynamic transition}, numerical simulations are performed to
estimate the efficiency of our scheme in various values of correlation
strengths. Section \ref{Summary} concludes this paper. Some nonessential
details of our calculation are placed in Appendixes.

\section{Model and $\protect\eta $ operators}

\label{model}

We consider a composite non-Hermitian system, described by the
Hamiltonian%
\begin{equation}
H=H_{\mathrm{A}}+H_{\mathrm{B}}+H_{\mathrm{AB}},
\end{equation}%
with Hermitian terms%
\begin{eqnarray}
H_{\mathrm{A}} &=&\frac{U}{2}\sum_{i=1}^{N_{a}}(a_{i,\uparrow }^{\dag
}a_{i,\uparrow }+a_{i,\downarrow }^{\dag }a_{i,\downarrow }), \\
H_{\mathrm{B}} &=&\sum_{\sigma =\uparrow ,\downarrow
}\sum_{i,j=1}^{N_{b}}(J_{ij}b_{i,\sigma }^{\dag }b_{j,\sigma }+\mathrm{H.c.}%
)+U\sum_{i=1}^{N_{b}}b_{i,\uparrow }^{\dag }b_{i,\downarrow }^{\dag
}b_{i,\downarrow }b_{i,\uparrow },  \notag
\end{eqnarray}%
and the non-Hermitian term%
\begin{equation}
H_{\mathrm{AB}}=\sum_{\sigma =\uparrow ,\downarrow
}\sum_{i=1}^{N_{a}}\sum_{j=1}^{N_{b}}\kappa _{ij}b_{j,\sigma }^{\dag
}a_{i,\sigma },
\end{equation}%
where $a_{i,\sigma }$\ and $b_{j,\sigma }$\ are fermion operators with spin$-%
\frac{1}{2}$ polarization $\sigma =\uparrow ,\downarrow $\ in lattices $%
N_{a} $\ and $N_{b}$, respectively. The parameters $J_{ij}$ ($i,j\in
N_{b},i\neq j$) and $\kappa _{ij}$\ ($i\in N_{a},j\in N_{b}$) are intra- and
inter-cluster hopping strengths, and taken to be real in this paper. Here
both $H_{\mathrm{A}}$\ and $H_{\mathrm{B}}$ are Hermitian, describing the
source system and the central system, respectively. $H_{\mathrm{A}}$\ is an
interaction-free system with trivial full flat band, while $H_{\mathrm{B}}$
is a standard Hubbard model, which is restricted to be the bipartite
lattice. In particular, the key features of the setup are (i) $H_{\mathrm{AB}%
}$\ is non-Hermitian,\ representing unidirectional tunnelings between two
subsystems\ $H_{\mathrm{A}}$\ and $H_{\mathrm{B}}$. (ii) The on-site
potential of a pair of fermions in $H_{\mathrm{A}}$ is identical to the
on-site repulsion in $H_{\mathrm{B}}$, but a little difference
will not affect the scheme since the EP dynamics can be extended to the
near-EP dynamics \cite{YXMPRB}. The schematic of the system is presented in
Fig. \ref{fig1}.

We define two $\eta $ operators for two subsystems

\begin{eqnarray}
\eta _{\mathrm{A}}^{\dag } &=&\sum_{i=1}^{N_{a}}\eta _{\mathrm{A},i}^{\dag
}=\sum_{i=1}^{N_{a}}\alpha _{i}a_{i,\uparrow }^{\dag }a_{i,\downarrow
}^{\dag }, \\
\eta _{\mathrm{B}}^{\dag } &=&\sum_{i=1}^{N_{b}}\eta _{\mathrm{B},i}^{\dag
}=\sum_{i=1}^{N_{b}}\beta _{i}b_{i,\uparrow }^{\dag }b_{i,\downarrow }^{\dag
},
\end{eqnarray}%
where $\alpha _{i}=\pm 1$ can be taken arbitrarily, since there are no
tunnelings between any two sites in the subsystem $A$, while $\beta _{i}=1$
and $-1$, for the different sublattice $i$ belongs to in the bipartite
lattice $B$. It can be shown that both operators satisfy%
\begin{equation}
\left[ H_{\mathrm{A}},\eta _{\mathrm{A}}^{\dag }\right] =U\eta _{\mathrm{A}%
}^{\dag },\left[ H_{\mathrm{B}},\eta _{\mathrm{B}}^{\dag }\right] =U\eta _{%
\mathrm{B}}^{\dag },
\end{equation}%
which can be utilized to construct the eigenstates of $H_{\mathrm{A}}$, $H_{%
\mathrm{B}}$, $H_{\mathrm{A}}+H_{\mathrm{B}}$,%
\begin{eqnarray}
\left\vert n\right\rangle _{\mathrm{A}} &=&\frac{1}{\sqrt{\Omega _{\mathrm{A}%
,n}}}\left( \eta _{\mathrm{A}}^{\dag }\right) ^{n}\left\vert \text{Vac}%
\right\rangle , \\
\left\vert m\right\rangle _{\mathrm{B}} &=&\frac{1}{\sqrt{\Omega _{\mathrm{B}%
,m}}}\left( \eta _{\mathrm{B}}^{\dag }\right) ^{m}\left\vert \text{Vac}%
\right\rangle ,
\end{eqnarray}%
where $\Omega _{\mathrm{A},n}=\left( n!\right) ^{2}C_{N_{a}}^{n}$\ and $%
\Omega _{\mathrm{B},m}=\left( m!\right) ^{2}C_{N_{b}}^{m}$ are normalization
factors.%
\begin{equation}
H_{\mathrm{A}}\left\vert n\right\rangle _{\mathrm{A}}=nU\left\vert
n\right\rangle _{\mathrm{A}},H_{\mathrm{B}}\left\vert m\right\rangle _{%
\mathrm{B}}=mU\left\vert m\right\rangle _{\mathrm{B}},
\end{equation}%
and%
\begin{equation}
\left( H_{\mathrm{A}}+H_{\mathrm{B}}\right) \left\vert n\right\rangle _{%
\mathrm{A}}\left\vert m\right\rangle _{\mathrm{B}}=\left( n+m\right)
U\left\vert n\right\rangle _{\mathrm{A}}\left\vert m\right\rangle _{\mathrm{B%
}}.
\end{equation}%
We can find that the set of eigenstates $\left\vert n\right\rangle _{\mathrm{%
A}}\left\vert m\right\rangle _{\mathrm{B}}$ are degenerate for fixed $m+n$.

In general, an $\eta $-pairing state can be regarded as the condensation of
bound pair fermions as hardcore boson. However, state $\left\vert
n\right\rangle _{\mathrm{A}}$ is trivial since it is just one of multi-fold
degenerate eigenstates. In addition, fully filled state $\left\vert
N_{a}\right\rangle _{\mathrm{A}}$\ and $\left\vert N_{b}\right\rangle _{%
\mathrm{B}}$\ are insulating states and can be easily prepared. The
desirable states are $\left\vert N_{a}\right\rangle _{\mathrm{A}}\left\vert
m\right\rangle _{\mathrm{B}}$ and $\left\vert 0\right\rangle _{\mathrm{A}%
}\left\vert m\right\rangle _{\mathrm{B}}$ with $1<m<N_{b}$, since both two
states possess ODLRO in the subsystem B.\

\section{Doublon effective Hamiltonian}

\label{Doublon effective Hamiltonian}

Like most interacting many-body systems, the exact solution of $H$\ is rare
although states $\left\vert n\right\rangle _{\mathrm{A}}\left\vert
m\right\rangle _{\mathrm{B}}$\ are eigenstates of $H_{\mathrm{A}}+H_{\mathrm{%
B}}$. In order to capture the physics of our scheme, we will consider the
problem in an energy shell. In the Hermitian system, one can employ the
perturbation method to get the effective Hamiltonian. However, the
corresponding theory has not been well established for the non-Hermitian
system, especially for the unidirectional hopping perturbation.%
In two Appendixes, we have illustrated how to obtain the effective
Hamiltonian of a non-Hermitian system from two perspectives. The Appendix A
provides an accurate effective Hamiltonian of a two-site non-Hermitian
system from the time evolution operator, while the Appendix B obtains the
effective Hamiltonian of an arbitrary-sized non-Hermitian system {for large $%
U$ limit }from parameters approaching the EP.

In this work, our aim is the dynamics for a special initial state, with the
subsystem A being fully occupied. It motivates us to consider pure doublon
states in the subsystem B, which has the same energy shell with that of the
initial state.  We take the parameter $\kappa _{ij}$ as a
constant {$\kappa \delta _{ij}$} for simplicity. For a subspace spanned by a
set of basis of doublon state $\left\{ {\left\vert \Psi _{\mathrm{D}}^{%
\mathrm{B}}\left( n\right) \right\rangle ,{n\in \lbrack 1,N_{b}]}}\right\} $,%
{\ the effective Hamiltonian can be written as} {%
\begin{equation}
H_{\mathrm{B}}^{\mathrm{eff}}=\sum_{i,j=1}^{N_{b}}\frac{-4J_{ij}^{2}}{U}(%
\boldsymbol{\eta }_{\mathrm{B},i}\cdot \boldsymbol{\eta }_{\mathrm{B},j}-%
\frac{1}{4})+U\sum_{i=1}^{N_{b}}\left( \frac{1}{2}+\eta _{\mathrm{B}%
,i}^{z}\right) ,
\end{equation}%
}in the case of $U\gg \left\vert J_{ij}\right\vert $. Here a doublon state
is {%
\begin{equation}
{\left\vert \Psi _{\mathrm{D}}^{\mathrm{B}}\left( n\right) \right\rangle =}%
b_{j_{1},\uparrow }^{\dag }b_{j_{1},\downarrow }^{\dag }b_{j_{2},\uparrow
}^{\dag }b_{j_{2},\downarrow }^{\dag }...b_{j_{n},\uparrow }^{\dag
}b_{j_{n},\downarrow }^{\dag }\left\vert \text{Vac}\right\rangle ,
\end{equation}%
with{\ $j_{n}\in \lbrack 1,N_{b}]$}, }and the pseudo-spin operator is
defined as $\boldsymbol{\eta }_{\mathrm{B},j}=(\eta _{\mathrm{B}%
,j}^{+}/2+\eta _{\mathrm{B},j}^{-}/2,\eta _{\mathrm{B},j}^{+}/2i-\eta _{%
\mathrm{B},j}^{-}/2i,\eta _{\mathrm{B},j}^{z})$ with $\eta _{\mathrm{B}%
,j}^{+}=\beta _{j}b_{j,\uparrow }^{\dag }b_{j,\downarrow }^{\dag }$ and $%
\eta _{\mathrm{B},j}^{z}=\left( n_{\mathrm{B},j,\uparrow }+n_{\mathrm{B}%
,j,\downarrow }-1\right) /2$.\  Similarly, for a subspace
spanned by a set of basis of doublon $\left\{ {\left\vert \Psi _{\mathrm{D}%
}^{\mathrm{A}}\left( n\right) \right\rangle ,{n\in \lbrack 1,N_{a}]}}%
\right\} $, which means ${n}${\ lattice sites in subsystem A occupied by two
particles with opposite }spin orientation, the effective Hamiltonian can be
written as%
\begin{equation}
H_{\mathrm{A}}^{\mathrm{eff}}=U\sum_{i=1}^{N_{a}}\left( \frac{1}{2}+\eta _{%
\mathrm{A},i}^{z}\right) ,
\end{equation}%
and corresponding operators obey the Lie algebra, i.e., $[\eta _{\mathrm{A}%
,i}^{+},$ $\eta _{\mathrm{A},j}^{-}]=2\eta _{\mathrm{A},j}^{z}\delta _{ij}$
and $[\eta _{\mathrm{A},i}^{z},$ $\eta _{\mathrm{A},j}^{\pm }]=\pm \eta _{%
\mathrm{A},j}^{\pm }\delta _{ij}$.

Now, it turns to establish the effective Hamiltonian of the non-Hermitian
term $H_{\mathrm{AB}}$. Unlike the Hermitian term, there is no unquestioned
perturbation theory for the non-Hermitian perturbation, especially near the
EP. In this work, we present the effective Hamiltonian $H_{\mathrm{AB}}^{%
\mathrm{eff}}$ from two perspectives. In two Appendixes, we
show that, for the given initial state with full filling A lattice and empty
B lattice, the dynamics obeys the effective Hamiltonian%
\begin{equation}
H^{\mathrm{eff}}=H_{\mathrm{A}}^{\mathrm{eff}}+H_{\mathrm{B}}^{\mathrm{eff}%
}+H_{\mathrm{AB}}^{\mathrm{eff}}
\end{equation}%
with%
\begin{equation}
H_{\mathrm{AB}}^{\mathrm{eff}}=\frac{4\kappa ^{2}}{U}\sum_{i}\eta _{\mathrm{%
A,}i}^{-}\eta _{\mathrm{B,}i}^{+},
\end{equation}%
where

\begin{equation}
\eta _{\mathrm{A},i}^{-}=\left( -1\right) ^{i}a_{i,\downarrow }a_{i,\uparrow
},\eta _{\mathrm{B},i}^{+}=\left( -1\right) ^{i}b_{i,\uparrow }^{\dag
}b_{i,\downarrow }^{\dag }.
\end{equation}%
It is clear that $H_{\mathrm{AB}}^{\mathrm{eff}}$\ is still a
non-Hermitian term which describes a unidirectional hopping of a doublon or
magnon from the point of view of spin wave.

Defining a total pseudo-spin operator%
\begin{equation}
\eta ^{z}=\sum_{i=1}^{N_{a}}\eta _{\mathrm{A},i}^{z}+\sum_{j=1}^{N_{b}}\eta
_{\mathrm{B},j}^{z},
\end{equation}%
we note that $\eta ^{z}$\ is conservative for the Hamiltonian $H^{\mathrm{eff%
}}$ due to the commutation relation%
\begin{equation}
\left[ \eta ^{z},H^{\mathrm{eff}}\right] =0,
\end{equation}%
which ensures that the Hilbert space of $H^{\mathrm{eff}}$\ can be
decomposed into many invariant subspaces labeled by the eigenvalues of $\eta
^{z}$, i.e., $2\eta ^{z}=-N_{a}-N_{b}$,\ $-N_{a}-N_{b}+1$,...,\ $%
N_{a}+N_{b}-1$,\ $N_{a}+N_{b}$. In this work, we only focus on the subspace
with $\eta ^{z}=\left( N_{a}-N_{b}\right) /2$ ($N_{a}<N_{b}$),\ which
contains the initial state with fully filling A sublattices and empty B
sublattices, i.e., $\prod_{i=1}^{N_{a}}a_{i,\uparrow }^{\dag
}a_{i,\downarrow }^{\dag }\left\vert \text{Vac}\right\rangle $.

\section{Jordan form with high-order EP}

\label{Jordan form with high-order EP}

In the above, we know that there are many degenerate eigenstates for $H_{%
\mathrm{A}}+H_{\mathrm{B}}$, which may become coalescing states when the
proper non-Hermitian term is added \cite{WPARXIV}. For non-Hermitian
operators, when the EP appears, there are eigenstates coalesce into one
state, leading to an incomplete Hilbert space \cite%
{Berry2004,Heiss2012,Miri2019,Zhang2020}. Mathematically, it relates to the
Jordan block form in the matrix \cite{Kato,Muller,Moiseyev,Emil}.
Remarkably, the peculiar features around the EP have sparked tremendous
attention to the classical and quantum photonic systems. The corresponding
intriguing dynamical phenomena include asymmetric mode switching \cite%
{Doppler2016}, topological energy transfer \cite{Xu2016}, robust wireless
power transfer \cite{Assawaworrarit2017}, and enhanced sensitivity \cite%
{Wiersig2014,Wiersig2016,Hodaei2017,Chen2017} depending on their EP
degeneracies. Many works have been devoted to the formation of the EP and
corresponding topological characterization in both theoretical and
experimental aspects \cite{Ding2016,Xiao2019,Pan2019}. In this work, we
employ the EP dynamics to prepare states with ODLRO. We start with the
Jordan form with high-order EP.

Considering two degenerate eigenstates $\left\vert A\right\rangle $ and $%
\left\vert B\right\rangle $ of the Hermitian Hamiltonian $H_{\mathrm{A}}+H_{%
\mathrm{B}}$, where

\begin{eqnarray}
\left\vert A\right\rangle &=&\left\vert N_{a}\right\rangle _{\mathrm{A}%
}\left\vert 0\right\rangle _{\mathrm{B}}=\frac{1}{\sqrt{\Omega _{\mathrm{A}%
,N_{a}}}}\left( \eta _{\mathrm{A}}^{\dag }\right) ^{N_{a}}\left\vert \text{%
Vac}\right\rangle , \\
\left\vert B\right\rangle &=&\left\vert 0\right\rangle _{\mathrm{A}%
}\left\vert N_{a}\right\rangle _{\mathrm{B}}=\frac{1}{\sqrt{\Omega _{\mathrm{%
B},N_{a}}}}\left( \eta _{\mathrm{B}}^{\dag }\right) ^{N_{a}}\left\vert \text{%
Vac}\right\rangle ,
\end{eqnarray}%
we have%
\begin{equation}
H\left\vert B\right\rangle =N_{a}U\left\vert B\right\rangle ,H^{\dag
}\left\vert A\right\rangle =N_{a}U\left\vert A\right\rangle ,
\end{equation}%
due to the facts%
\begin{equation}
H_{\mathrm{AB}}\left\vert 0\right\rangle _{\mathrm{A}}\left\vert
N_{a}\right\rangle _{\mathrm{B}}=0,\left( H_{\mathrm{AB}}\right) ^{\dag
}\left\vert N_{a}\right\rangle _{\mathrm{A}}\left\vert 0\right\rangle _{%
\mathrm{B}}=0.
\end{equation}%
It means that two states $\left\vert A\right\rangle $ and $\left\vert
B\right\rangle $ are mutually biorthogonal conjugate and $\langle
A\left\vert B\right\rangle $\ is the biorthogonal norm of them. On the other
hand, we have
\begin{equation}
\langle A\left\vert B\right\rangle =0.
\end{equation}%
The vanishing norm indicates that state $\left\vert B\right\rangle $($%
\left\vert A\right\rangle $) is the coalescing state of $H$($H^{\dag }$), or
Hamiltonians $H$\ and $H^{\dag }$\ get an EP.

However, it is a little hard to determine the corresponding Jordan block
form and the order of the EP. In the following, we estimate the order in
large $U$ limit. At first, the above analysis for two states $\left\vert
A\right\rangle $ and $\left\vert B\right\rangle $ is applicable for the
effective Hamiltonian $H^{\mathrm{eff}}$. This means that there is an EP in
the invariant subspace with $\eta ^{z}=\left( N_{a}-N_{b}\right) /2$, and
dimension $C_{N_{a}+N_{b}}^{N_{a}}$. The order of such an EP is determined
by the corresponding Jordan block. Second, when we consider a complete set
of degenerate eigenstates of the Hermitian Hamiltonian $H_{\mathrm{A}}+H_{%
\mathrm{B}}$ in this subspace, which are denoted as $\left\{ \left\vert
n\right\rangle _{\mathrm{A}}\left\vert m\right\rangle _{\mathrm{B}}\right\} $
($n\in \left[ 0,N_{a}\right] ,N_{a}\leqslant N_{b}$) with fixed $m+n=N_{a}$,
the effective Hamiltonian can be expressed as an $\left( N_{a}+1\right)
\times \left( N_{a}+1\right) $ matrix $M$ with nonzero matrix elements

\begin{eqnarray}
&&\left( M\right) _{N_{a}+1-n,N_{a}-n}  \notag \\
&=&\left\langle N_{a}-n\right\vert _{\mathrm{B}}\left\langle n\right\vert _{%
\mathrm{A}}H^{\mathrm{eff}}\left\vert n+1\right\rangle _{\mathrm{A}%
}\left\vert N_{a}-n-1\right\rangle _{\mathrm{B}}  \notag \\
&=&\frac{4\kappa ^{2}}{U}\frac{N_{a}-n}{N_{b}}\sqrt{\left( n+1\right) \left(
N_{b}-N_{a}+n+1\right) }
\end{eqnarray}%
with $n=\left[ 0,N_{a}-1\right] $, and%
\begin{eqnarray}
&&\left( M\right) _{N_{a}+1-n,N_{a}+1-n}  \notag \\
&=&\left\langle N_{a}-n\right\vert _{\mathrm{B}}\left\langle n\right\vert _{%
\mathrm{A}}H^{\mathrm{eff}}\left\vert n\right\rangle _{\mathrm{A}}\left\vert
N_{a}-n\right\rangle _{\mathrm{B}}  \notag \\
&=&N_{a}U
\end{eqnarray}%
with $n=\left[ 0,N_{a}\right] $. It is obviously an $\left( N_{a}+1\right) $%
-order Jordan block, satisfying
\begin{eqnarray}
&&\left[ \left( M-N_{a}UI\right) ^{N_{a}}\right] _{ij}=\prod%
\limits_{n=0}^{N_{a}-1}\frac{4\kappa ^{2}}{U}\frac{N_{a}-n}{N_{b}}  \notag \\
&&\times \sqrt{\left( n+1\right) \left( N_{b}-N_{a}+n+1\right) }\delta
_{N_{a}+1,1}.
\end{eqnarray}%
where $I$\ is the unit matrix. In other words, matrix $\left(M-N_{a}UI%
\right) $ is a nilpotent matrix, i.e.,%
\begin{equation}
\left( M-N_{a}UI\right) ^{N_{a}+1}=0.
\end{equation}%
Taking $N_{a}=N_{b}=4$, for example, the matrix has the form%
\begin{equation}
M=\frac{2\kappa ^{2}}{U}\left(
\begin{array}{ccccc}
0 & 0 & 0 & 0 & 0 \\
2 & 0 & 0 & 0 & 0 \\
0 & 3 & 0 & 0 & 0 \\
0 & 0 & 3 & 0 & 0 \\
0 & 0 & 0 & 2 & 0%
\end{array}%
\right) +4UI,
\end{equation}%
which possesses a single eigenvector $\left(
\begin{array}{ccccc}
0 & 0 & 0 & 0 & 1%
\end{array}%
\right) ^{T}$.

\begin{figure*}[tbp]
\includegraphics[ bb=9 83  610 730, width=0.25\textwidth, clip]{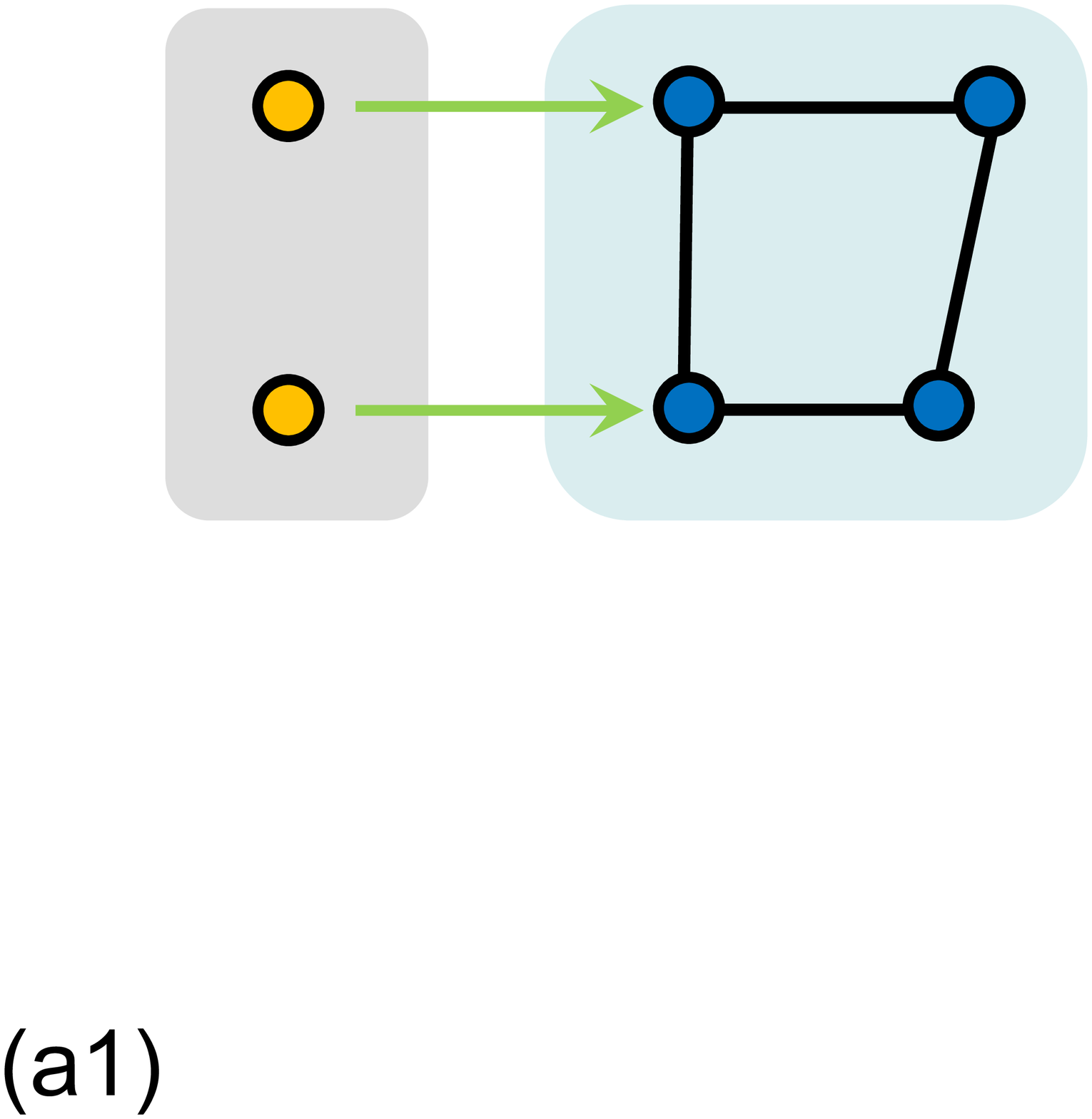} %
\includegraphics[ bb=0 0 390 303, width=0.35\textwidth, clip]{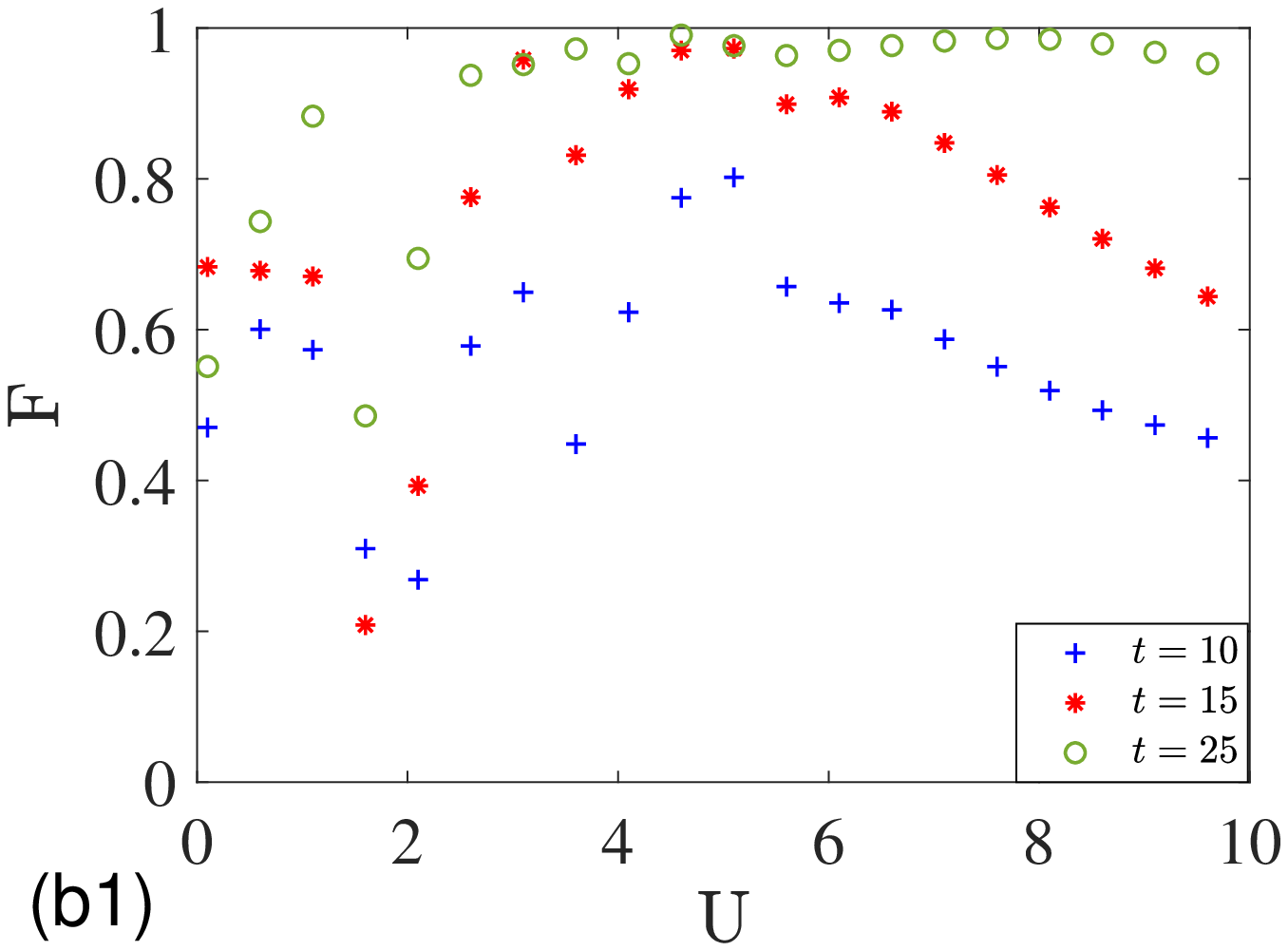} %
\includegraphics[ bb=0 0 390 303, width=0.348\textwidth, clip]{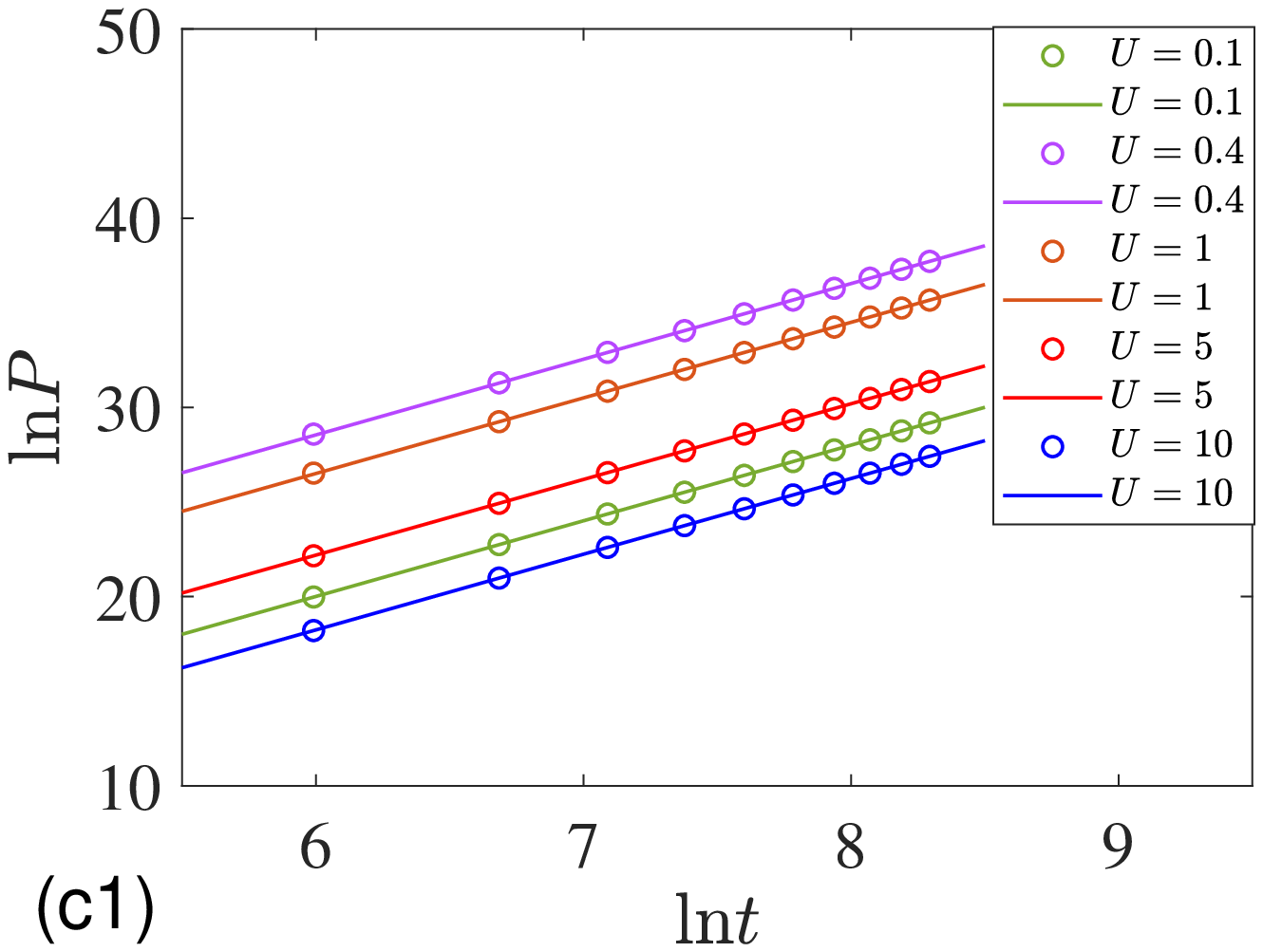} %
\includegraphics[ bb=9 83  610 730, width=0.25\textwidth, clip]{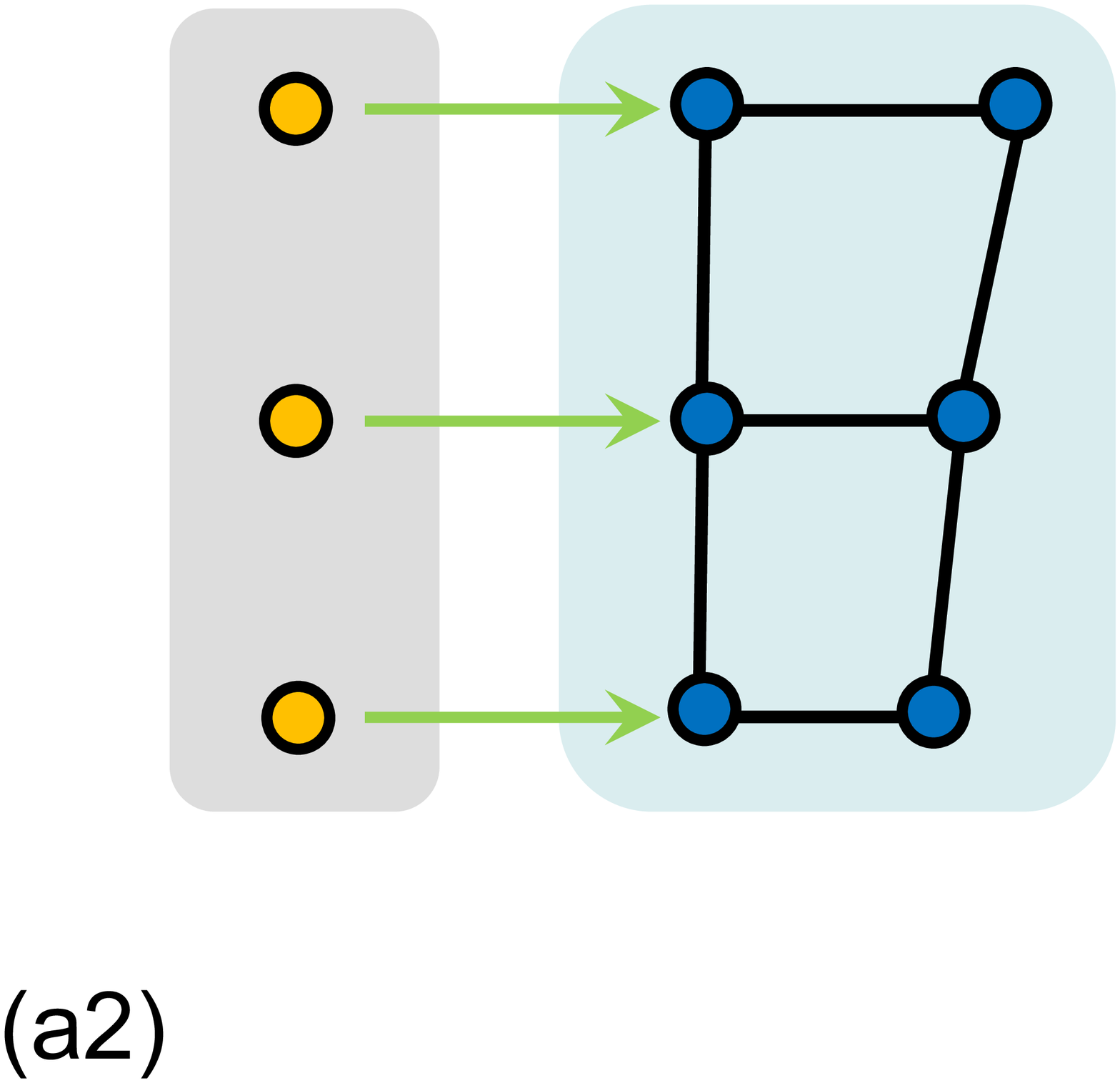} %
\includegraphics[ bb=0 0 390 303, width=0.35\textwidth, clip]{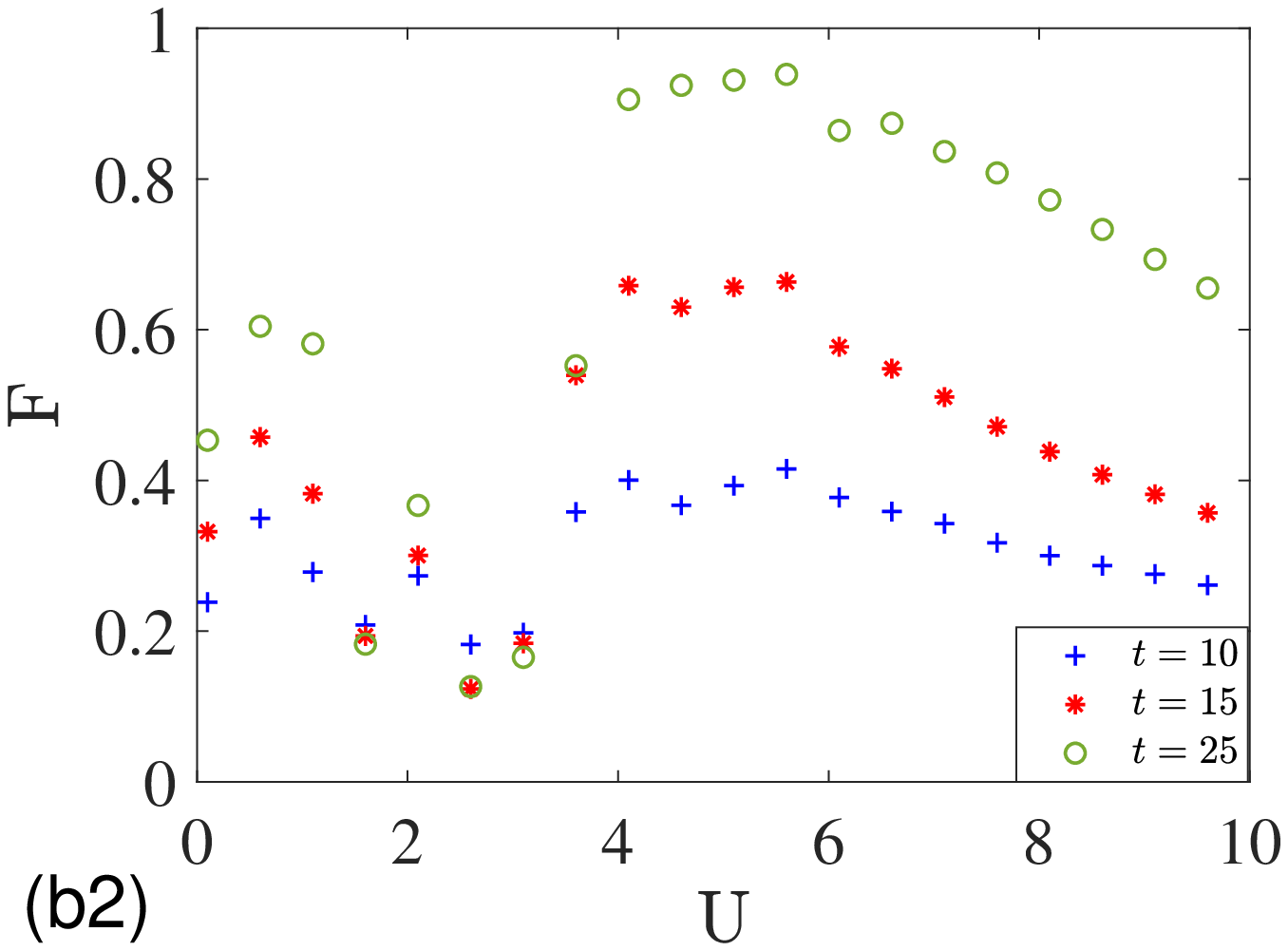} %
\includegraphics[ bb=0 0 390 303, width=0.348\textwidth, clip]{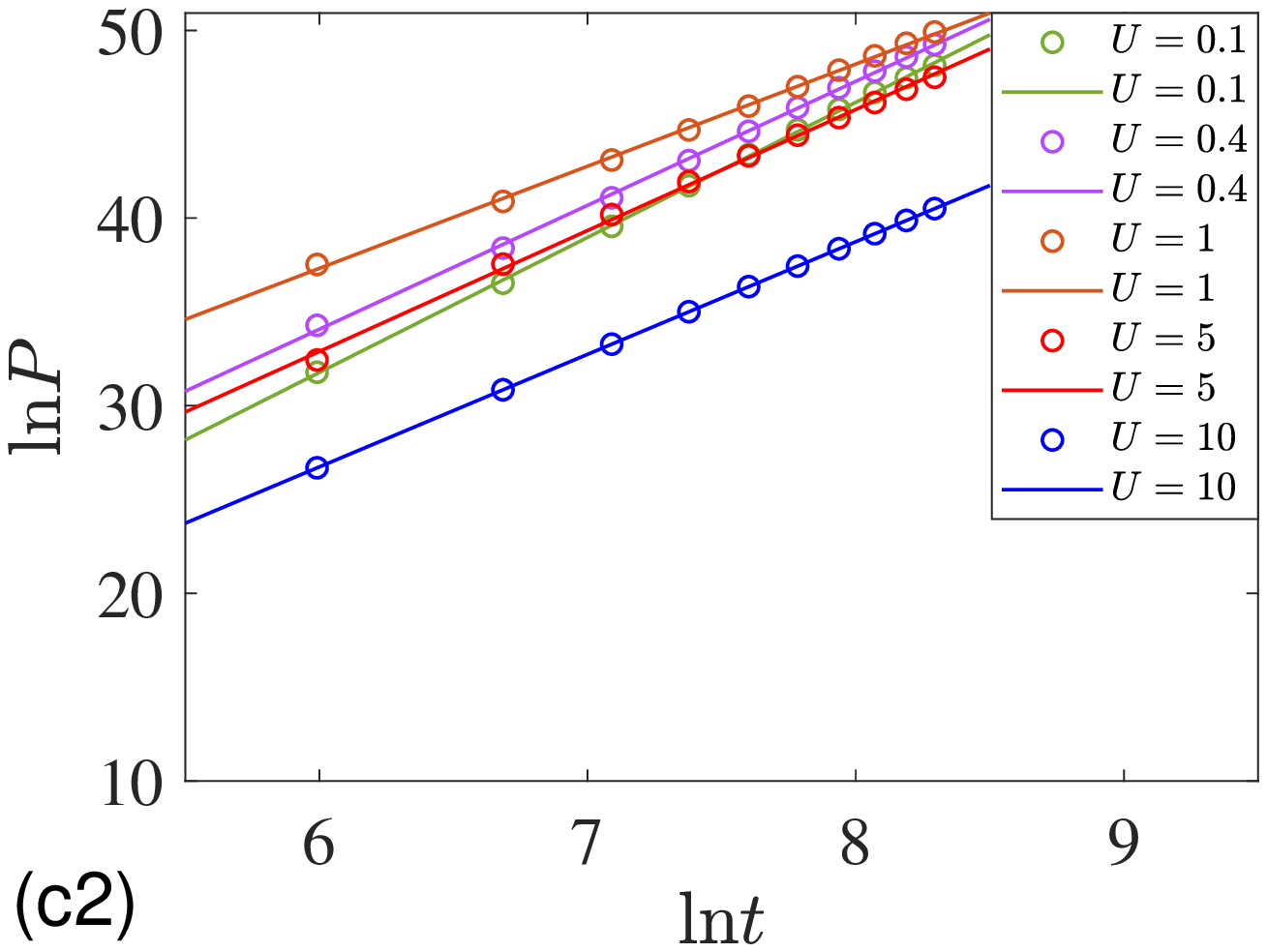}
\caption{Schematic illustration of (a1) 6 sites composite %
non-Hermitian system with 4 filled particles and (a2) 9 sites composite non-Hermitian system with 6 filled particles concerned in the
following numerical simulations. The parameters are $\protect\kappa =1$,
(a1)-(c1) $N_{a}=2$, $N_{b}=4$, $J_{ij}=0.75$, $1.17$, $0.68$, $1.02$;
(a2)-(c2) $N_{a}=3$, $N_{b}=6$, $J_{ij}=0.75$, $1.17$, $0.68$, $1.02$, $0.87$%
, $0.61$, $0.72$. (b1) and (b2) are plots of fidelity $F(t)$ defined in Eq. (%
\protect\ref{fidelity}). Three typical values of $t$ are taken and indicated
in the panels. The numerical data oscillate at small $U$ values and the
optimal $U$ occurs around $U=5$. (c1) and (c2) show the scaling law of
probability $P$ defined in Eq. (\protect\ref{P}) as a function of $t$ for
different value of $U$. Different colored dots represent the numerical date,
which are fitted by different colored solid lines (c1) $\ln P=3.99\ln t-3.96$%
, $\ln P=3.97\ln t+4.77$, $\ln P=3.98\ln t+2.67$, $\ln P=3.99\ln t-1.81$, $%
\ln P=4.00\ln t-5.76$ and (c2) $\ln P=7.20\ln t-11.40$, $\ln P=6.61\ln
t-5.56 $, $\ln P=5.44\ln t+4.66$, $\ln P=6.45\ln t-5.80$, $\ln P=6.00\ln
t-9.26$ from $U=0.1$ to $U=10$ respectively.}
\label{fig2}
\end{figure*}

The dynamics for any states in this subspace is governed by the time
evolution operator%
\begin{equation}
U(t)=e^{-iMt}=e^{-iN_{a}Ut}\sum_{l=0}^{N_{a}}\frac{1}{l!}\left[ -i\left(
M-N_{a}UI\right) t\right] ^{l}.
\end{equation}%
It indicates that for the initial state $\left\vert \Psi (0)\right\rangle
=\left\vert A\right\rangle $, we have%
\begin{eqnarray}
\left\vert \Psi (t)\right\rangle &=&e^{-iMt}\left\vert A\right\rangle \\
&=&e^{-iN_{a}Ut}\left(
\begin{array}{cccccc}
1 & f_{1} & ... & f_{q} & ... & f_{N_{a}}%
\end{array}%
\right) ^{T},  \notag
\end{eqnarray}%
where the elements%
\begin{eqnarray}
f_{q} &=&\sqrt{A_{N_{a}}^{q}A_{N_{b}}^{q}}\left( \frac{-4i\kappa ^{2}t}{%
UN_{b}}\right) ^{q},q\in \lbrack 1,N_{a}] \\
A_{N_{a}}^{q} &=&\frac{N_{a}!}{\left( N_{a}-q\right) !},A_{N_{b}}^{q}=\frac{%
N_{b}!}{\left( N_{b}-q\right) !},
\end{eqnarray}%
and
\begin{equation}
\left\vert \left\vert \Psi \left( t\right) \right\rangle \right\vert =\sqrt{%
1+\sum\limits_{q=1}^{N_{a}}\left\vert f_{q}\right\vert ^{2}}\approx
\left\vert f_{N_{a}}\right\vert \text{ }
\end{equation}%
at large $t\gg \frac{UN_{b}}{\kappa ^{2}}$. Setting the target state as%
\begin{equation}
\left\vert \Psi _{\text{target}}\right\rangle =\left\vert B\right\rangle ,
\end{equation}%
we have the fidelity%
\begin{equation}
F(t)=\frac{\left\vert \left\langle \Psi _{\text{target}}\right\vert \Psi
\left( t\right) \rangle \right\vert }{\left\vert \left\vert \Psi \left(
t\right) \right\rangle \right\vert }=\frac{\left\vert f_{N_{a}}\right\vert }{%
\left\vert \left\vert \Psi \left( t\right) \right\rangle \right\vert }%
\approx 1,
\end{equation}%
which indicates that state $\left\vert \Psi _{\text{target}}\right\rangle $
becomes dominant in the evolved state $\left\vert \Psi \left( t\right)
\right\rangle $\ at large $t\gg \frac{UN_{b}}{\kappa ^{2}}$. The increasing
behavior of $\left\vert \left\vert \Psi \left( t\right) \right\rangle
\right\vert $ obeying $\left\vert \left\vert \Psi \left( t\right)
\right\rangle \right\vert ^{2}\propto t^{2N_{a}}$\ within large $t$\ region
is also a dynamic demonstration for the order of the Jordan block. This
analytical analysis shows that the speed of relaxation of ODLRO pair state
depends on the order of the EP, which is determined by the number of pairs.
Furthermore, we would like to address two points. (i) The non-Hermitian
effective Hamiltonian $H_{\mathrm{AB}}^{\mathrm{eff}}$\ is obtained from the
simplest case with $N_{a}=N_{b}=1$\ in the Appendix A. Its validity for %
large systems is illustrated in the Appendix B for large $U$
limit and from the perspective of parameters approaching the EP. (ii) The
power behavior of $\left\vert \left\vert \Psi \left( t\right) \right\rangle
\right\vert $\ requires large $t$.\ However, in practice, $F(t)$\ may
approach unit before this time domain.

\section{Dynamic transition}

\label{Dynamic transition}

The above analysis provides a prediction about the dynamic transition from
an insulating state to an $\eta $-pairing state in a composite %
non-Hermitian system. The composite system is consisted of two parts (or
two layers), one is a trivial system (source system) constructed by a set of
isolated sites, while the other is a Hubbard model (central system), which
supports $\eta $-pairing eigenstates. Initially, two subsystems are
separated and the source system is fully filled by electrons, being in an
insulating state, while the central system is empty. The decoupling between
two subsystems can be achieved in two ways, i.e., the pre-quench Hamiltonian
can be setted by (a) taking the chemical potential on source system far from
the resonant energy of the central system; (b) switching off the tunneling
terms between two subsystems directly under the resonant condition. The\
post-quench Hamiltonian is then $H_{\mathrm{A}}+H_{\mathrm{B}}+H_{\mathrm{AB}%
}$. According to our analysis, both two quench dynamics should result in
steady\ superconducting state, realizing the dynamic transition from an
insulating state to an $\eta $-pairing state.

\begin{figure}[tbp]
\includegraphics[ bb=0 0 500 410, width=0.45\textwidth, clip]{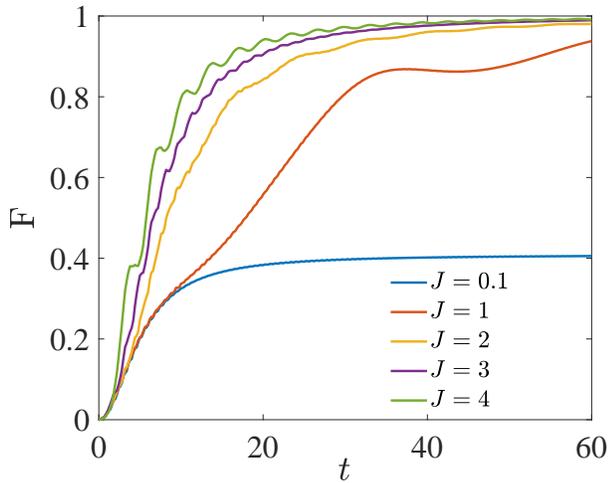}
\caption{The Plot of $F(t)$, which is obtained by exact
diagonalization of the original Hamiltonian for the finite system. The
parameters are $N_{a}=2$, $N_{b}=4$, $\protect\kappa =1$, $U=20$, $J_{ij}=J$%
. Several typical values of $J$ are taken and indicated in the panels.}
\label{figR1}
\end{figure}

\begin{figure*}[tbp]
\includegraphics[ bb=40 545 343 790, width=0.32\textwidth, clip]{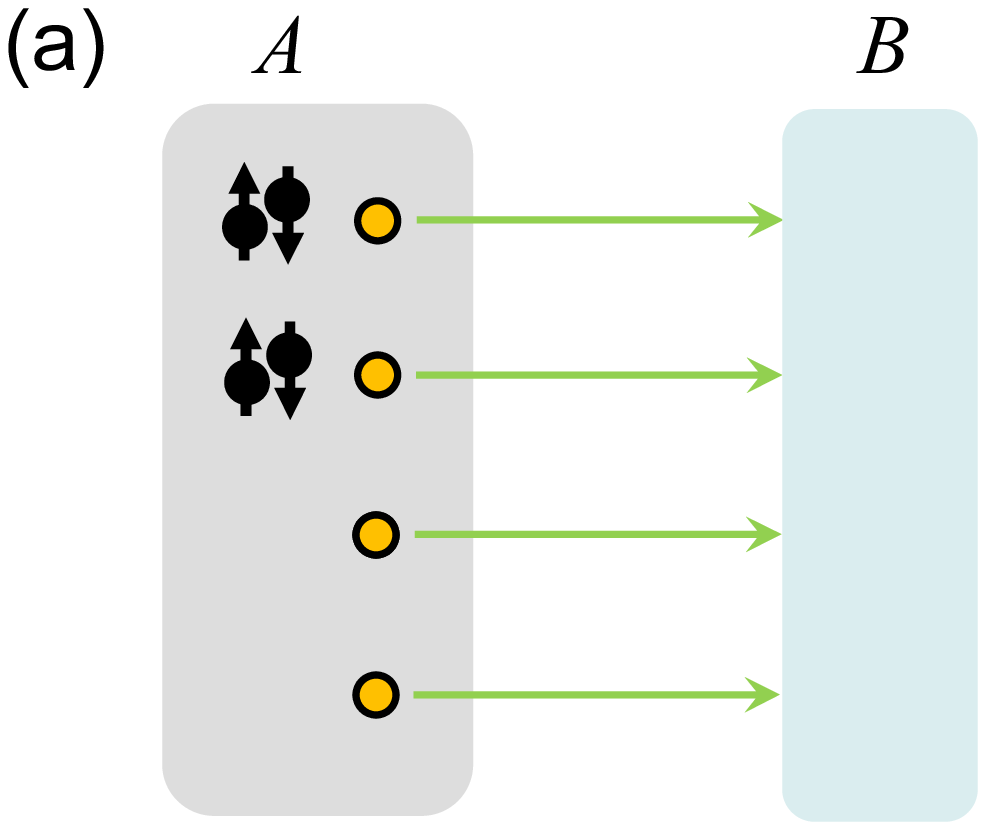} %
\includegraphics[ bb=40 545 343 790, width=0.32\textwidth, clip]{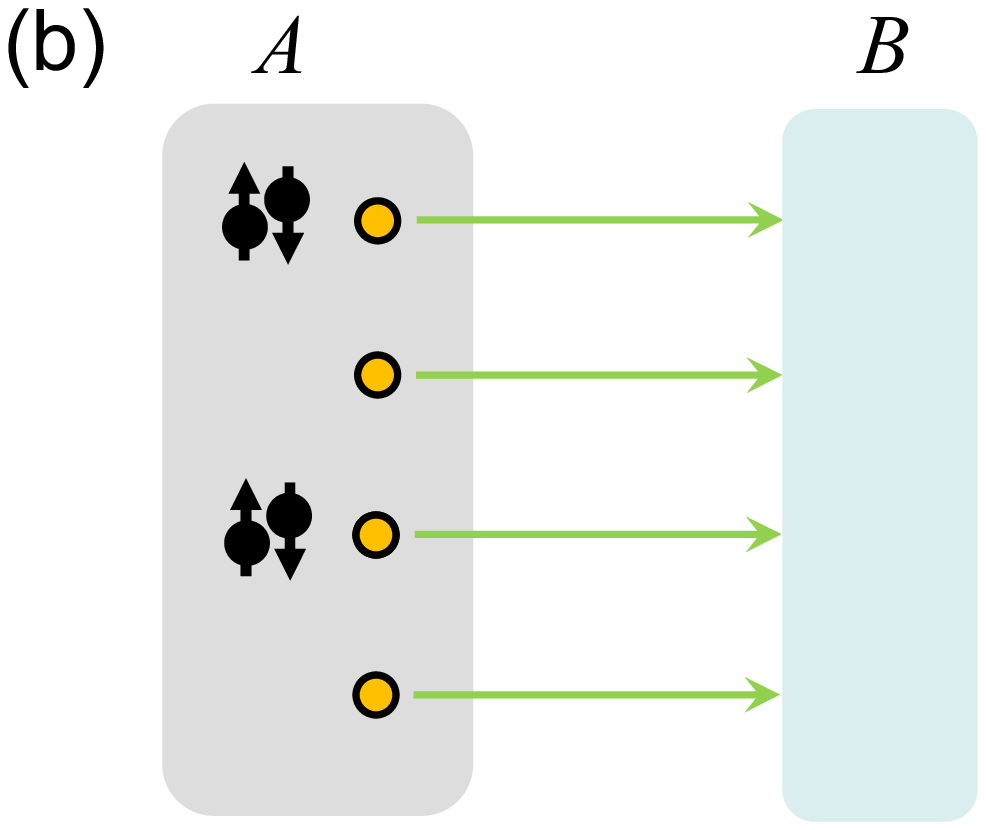} %
\includegraphics[ bb=40 545 343 790, width=0.32\textwidth, clip]{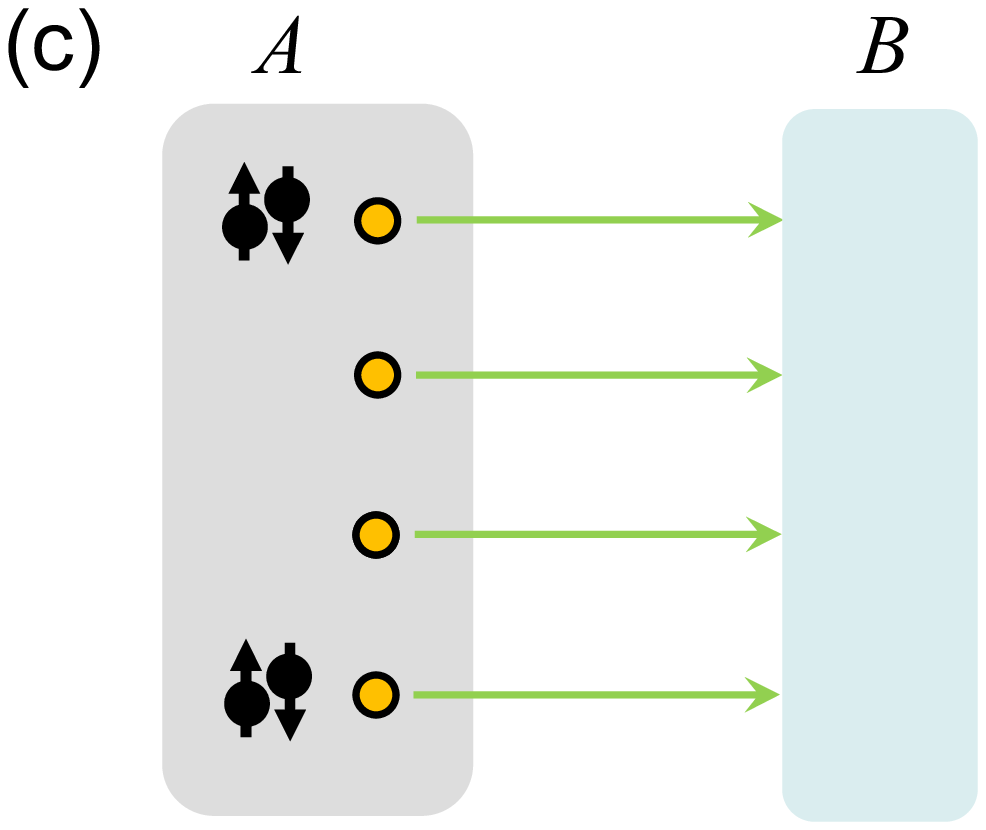}
\caption{Schematics of the composite non-Hermitian system
with different initial states. (a), (b), and (c) present three typical
configurations, in which two doublons are located in different sites of
system A. The analytical analysis based on the perturbation method in large $%
U$ limit indicates that the final states after long time are the same.
Numerical simulations for finite $U$ and small-size system support this
conclusion.}
\label{fig3}
\end{figure*}

{We perform numerical simulations on finite system with the following
considerations. (i) The analysis based on the effective
Hamiltonian in last section only predicts the results for large }${U}${\
within large time domain. The efficiency of the scheme should be estimated
from numerical simulations of the original Hamiltonian}. (ii)
The existence of $\eta $-pairing{\ eigenstates are independent of the
distribution of the hoppings for B sublattices. }The evolved states $%
\left\vert \Psi \left( t\right) \right\rangle $\ for initial states $%
\left\vert 2\right\rangle _{\mathrm{A}}\left\vert 0\right\rangle _{\mathrm{B}%
}$\ and $\left\vert 3\right\rangle _{\mathrm{A}}\left\vert 0\right\rangle _{%
\mathrm{B}}$\ in two finite systems are computed by exact diagonalization.
We focus on the Dirac probability%
\begin{equation}
P(t)=\left\vert \left\vert \Psi \left( t\right) \right\rangle \right\vert
^{2},  \label{P}
\end{equation}%
and the fidelity%
\begin{equation}
F(t)=\frac{1}{\sqrt{P}}\left\vert \left\langle \Psi _{\text{target}%
}\right\vert \Psi \left( t\right) \rangle \right\vert ,  \label{fidelity}
\end{equation}%
with the target states being $\left\vert 0\right\rangle _{\mathrm{A}%
}\left\vert 2\right\rangle _{\mathrm{B}}$\ and $\left\vert 0\right\rangle _{%
\mathrm{A}}\left\vert 3\right\rangle _{\mathrm{B}}$, respectively. {The
lattice geometry and numerical results are plotted in Fig. \ref{fig2}. We
plot the fidelity as function of }${U}${\ for three typical instants. We
find that there exits an optimal }${U\approx 5}${, at which the fidelity
gets the maximal value. We also plot the probability ln}${P(t)}${\ as
function of ln}${t}${\ to demonstrate the EP dynamic behavior. }From the
results of linear fitting, it can be seen that the slope of the line
deviates from the predicted value for the cases with small $U$, especially
for larger system. This indicates that the speed of relaxation of pair state
depends on the order of the EP and the fidelity of the scheme is immune to
the irregularity of the lattice. We estimated the relation of
the hopping strength and the efficiency of the scheme from numerical
simulations of the original Hamiltonian in Fig. \ref{figR1}. Within a
certain range of parameters, the increase of the hopping strength will
improve the efficiency of the scheme.

\section{Summary and discussion}

\label{Summary}

In summary, we have extended the scheme of quantum casting to interacting
many-body systems. Unlike the previous work Ref. \cite{YXMPRB} on
non-interacting systems, the present scheme does not require the scan on the
chemical potential of the source system. Our findings offer a method for the
efficient preparation of correlated states and are expected to be necessary
and insightful for quantum engineering. The key point is the exceptional
dynamics, which allows particles pumping from the source subsystem to the
central subsystem, realizing the dynamical preparation of many-body quantum
states. It is due to the resonance between the initial state and the target
state. Accordingly, there is a class of initial states (see {Fig. \ref{fig3}}%
) evolving to the same final state after long time. Numerical simulations
for finite $U$ and small-size system support this conclusion. In this sense,
such a scheme can be applied to other interacting many-body systems. On the
other hand, considering a quench process with the pre-quench Hamiltonian
being $H_{\mathrm{A}}+H_{\mathrm{B}}$, and {the\textbf{\ }}post-quench
Hamiltonian{\ being }$H_{\mathrm{A}}+H_{\mathrm{B}}+H_{\mathrm{AB}}${, the
Loschmidt echo }$\left\vert L\right\vert ^{2}=\left\vert \left\langle \Psi
\left( 0\right) \right\vert \Psi \left( t\right) \rangle \right\vert ^{2}${\
should turn to zero after a long time for a finite system. It may predict an
\textit{asymptotic} dynamic quantum phase transition (DQPT) \cite{QD3} in
thermodynamic limit, i.e., }$\left\vert L\right\vert ^{2}${\ decays rapidly,
rather than vanishes at a finite instant in a standard DQPT. The final
answer depends on }the scaling behavior of $\left\vert L\right\vert ^{2}$,
that is an open question at the present stage.

\acknowledgments This work was supported by National Natural Science
Foundation of China (under Grant No. 11874225).

\section*{Appendix A}

\label{Appendix1}

In this Appendix, we present a derivation of the effective Hamiltonian in
the doublon subspace for the tunneling term between two subsystems A and B.
We will obtain the effective Hamiltonian from the time evolution operator
rather than the perturbation method due to the concern with the availability
of it for a non-Hermitian system at exceptional point.

Consider a two-site Hamiltonian

\begin{equation}
H_{\mathrm{conn}}=\sum_{\sigma =\uparrow ,\downarrow }\kappa b_{\sigma
}^{\dag }a_{\sigma }+Ub_{\uparrow }^{\dag }b_{\downarrow }^{\dag
}b_{\downarrow }b_{\uparrow }+\frac{U}{2}(a_{\uparrow }^{\dag }a_{\uparrow
}+a_{\downarrow }^{\dag }a_{\downarrow }),
\end{equation}%
which describes the connection between any two sites among A and B
subsystems. We neglect the subscripts of the operators for the sake of
simplicity. We start from the matrix representation of the Hamiltonian in
the invariant subspace spanned by the basis set

\begin{eqnarray}
\left\vert 1\right\rangle &=&\left\vert \uparrow \downarrow \right\rangle _{%
\mathrm{A}}\left\vert 0\right\rangle _{\mathrm{B}}=a_{\uparrow }^{\dag
}a_{\downarrow }^{\dag }\left\vert \mathrm{Vac}\right\rangle \\
\left\vert 2\right\rangle &=&\left\vert 0\right\rangle _{\mathrm{A}%
}\left\vert \uparrow \downarrow \right\rangle _{\mathrm{B}}=b_{\uparrow
}^{\dag }b_{\downarrow }^{\dag }\left\vert \mathrm{Vac}\right\rangle  \notag
\\
\left\vert 3\right\rangle &=&\left\vert \uparrow \right\rangle _{\mathrm{A}%
}\left\vert \downarrow \right\rangle _{\mathrm{B}}=a_{\uparrow }^{\dag
}b_{\downarrow }^{\dag }\left\vert \mathrm{Vac}\right\rangle  \notag \\
\left\vert 4\right\rangle &=&\left\vert \downarrow \right\rangle _{\mathrm{A}%
}\left\vert \uparrow \right\rangle _{\mathrm{B}}=a_{\downarrow }^{\dag
}b_{\uparrow }^{\dag }\left\vert \mathrm{Vac}\right\rangle  \notag
\end{eqnarray}%
is%
\begin{equation}
h=\left(
\begin{array}{cccc}
U & 0 & 0 & 0 \\
0 & U & \kappa & -\kappa \\
\kappa & 0 & \frac{U}{2} & 0 \\
-\kappa & 0 & 0 & \frac{U}{2}%
\end{array}%
\right) ,
\end{equation}%
which contains a $2\times 2$\ Jordan block for nonzero $U$. The solution of
matrix consists $3$ eigenvectors $\left\vert \phi _{c}\right\rangle $, $%
\left\vert \phi _{3}\right\rangle $ and $\left\vert \phi _{4}\right\rangle $%
, with eigenvalues $U$, $U/2$, and $U/2$, respectively. The explicit form of
the vectors is
\begin{eqnarray}
\left\vert \phi _{a}\right\rangle &=&\left(
\begin{array}{c}
1 \\
0 \\
\frac{2\kappa }{U} \\
-\frac{2\kappa }{U}%
\end{array}%
\right) ,\left\vert \phi _{c}\right\rangle =\left(
\begin{array}{c}
0 \\
1 \\
0 \\
0%
\end{array}%
\right) , \\
\left\vert \phi _{3}\right\rangle &=&\left(
\begin{array}{c}
0 \\
-\frac{2\kappa }{U} \\
1 \\
0%
\end{array}%
\right) ,\left\vert \phi _{4}\right\rangle =\left(
\begin{array}{c}
0 \\
\frac{2\kappa }{U} \\
0 \\
1%
\end{array}%
\right) ,  \notag
\end{eqnarray}%
where $\left\vert \phi _{c}\right\rangle $ is the coalescing vector and $%
\left\vert \phi _{a}\right\rangle $\ is the corresponding auxiliary vector,
satisfying%
\begin{equation}
\left( h-UI\right) \left\vert \phi _{a}\right\rangle =\left\vert \phi
_{c}\right\rangle ,
\end{equation}%
where $I$ is the unit matrix. We would like to point out that in the case of
$U=0$, $h$ contains a $3\times 3$\ Jordan block.\textbf{\ }The solution of
matrix consists $2$ eigenvectors $\left\vert \phi _{c}\right\rangle $ and $%
\left\vert \phi _{4}\right\rangle $, with the same eigenvalue $0$. The
explicit form of the vectors is%
\begin{equation}
\left\vert \phi _{a}\right\rangle =\left(
\begin{array}{c}
1 \\
0 \\
0 \\
0%
\end{array}%
\right) ,\left\vert \phi _{c}\right\rangle =\left(
\begin{array}{c}
0 \\
1 \\
0 \\
0%
\end{array}%
\right) ,\left\vert \phi _{4}\right\rangle =\left(
\begin{array}{c}
0 \\
0 \\
1 \\
1%
\end{array}%
\right) ,
\end{equation}%
where $\left\vert \phi _{c}\right\rangle $ is the coalescing vector and $%
\left\vert \phi _{a}\right\rangle $\ is the corresponding auxiliary vector.
In this work, we only focus on the case with nonzero $U$. However, one
should consider the effect of $3$-order EP when $U$ is very small. Then the
time evolution operator in such a invariant subspace can be obtained as%
\begin{equation}
e^{-iht}=e^{-itU}\left(
\begin{array}{cccc}
1 & 0 & 0 & 0 \\
-\frac{4it\kappa ^{2}}{U}-\frac{8\kappa ^{2}}{U^{2}}\Lambda & 1 & \frac{%
2\kappa }{U}\Lambda & -\frac{2\kappa }{U}\Lambda \\
\frac{2\kappa }{U}\Lambda & 0 & e^{\frac{itU}{2}} & 0 \\
-\frac{2\kappa }{U}\Lambda & 0 & 0 & e^{\frac{itU}{2}}%
\end{array}%
\right) ,
\end{equation}%
where $\Lambda =1-e^{\frac{itU}{2}}$. The time evolution of the trivial
initial state $\left\vert \phi _{a}\right\rangle $ can be expressed as
\begin{equation}
\left\vert \Psi \left( t\right) \right\rangle =e^{-iht}\left\vert \phi
_{a}\right\rangle =e^{-itU}\left(
\begin{array}{c}
1 \\
-\frac{4it\kappa ^{2}}{U}-\frac{8\kappa ^{2}}{U^{2}}\Lambda \\
\frac{2\kappa }{U} \\
-\frac{2\kappa }{U}%
\end{array}%
\right) .
\end{equation}%
We find that
\begin{equation}
\langle 2\left\vert \Psi \left( t\right) \right\rangle =-e^{-itU}\frac{%
4\kappa ^{2}}{U}\left[ it+\frac{2}{U}\left( 1-e^{\frac{itU}{2}}\right) %
\right] ,
\end{equation}%
which can be valued within finite time scale as%
\begin{equation}
\langle 2\left\vert \Psi \left( t\right) \right\rangle \approx -e^{-itU}%
\frac{4\kappa ^{2}}{U}\left\{
\begin{array}{cc}
it, & \text{large }U \\
\frac{U}{4}t^{2}, & \text{small }U%
\end{array}%
\right. .
\end{equation}%
Note that the switching of powers of the variable $t$ is due to the
cancellation of the linear $t$ term in the small $U$\ limit. It accords with
the above analysis about the order of Jordan block. In large $U$ limit, $%
U\gg \kappa $, $e^{-iht}$ reduces to a diagonal-block form%
\begin{equation}
e^{-iht}\approx e^{-itU}\left(
\begin{array}{cccc}
1 & 0 & 0 & 0 \\
-\frac{4it\kappa ^{2}}{U} & 1 & 0 & 0 \\
0 & 0 & e^{\frac{itU}{2}} & 0 \\
0 & 0 & 0 & e^{\frac{itU}{2}}%
\end{array}%
\right) .
\end{equation}%
Then in the doublon subspace spanned by $a_{\uparrow }^{\dag }a_{\downarrow
}^{\dag }\left\vert \mathrm{Vac}\right\rangle $ and $b_{\uparrow }^{\dag
}b_{\downarrow }^{\dag }\left\vert \mathrm{Vac}\right\rangle $, the
effective Hamiltonian is%
\begin{equation}
H_{\mathrm{conn}}^{\text{eff}}=\frac{4\kappa ^{2}}{U}b_{\uparrow }^{\dag
}b_{\downarrow }^{\dag }a_{\downarrow }a_{\uparrow }+UI_{2}.
\end{equation}

\section*{Appendix B} \label{Appendix2} In this
appendix, we obtain the effective Hamiltonian of the the tunneling term $%
H_{AB}$ with arbitrary size for large $U$ limit from the perspective of
parameters approaching the EP. At first, we add an unidirectional tunneling
term $\lambda a_{l,\sigma }^{\dag }b_{l,\sigma }$ and take the parameter {$%
\kappa _{ij}$ as a constant $\kappa \delta _{ij}$ for simplicity}. The new
tunneling between two subsystems $H_{A}$ and $H_{B}$ reads  %
\begin{equation}
H_{\mathrm{AB}}^{\prime }=\sum_{\sigma =\uparrow ,\downarrow }\sum_{l}\left(
\kappa b_{l,\sigma }^{\dag }a_{l,\sigma }+\lambda a_{l,\sigma }^{\dag
}b_{l,\sigma }\right) .
\end{equation}%
We introduce a set of canonical operators
\begin{eqnarray}
\bar{c}_{l,\sigma } &=&\sqrt{\frac{\lambda }{\kappa }}a_{l,\sigma }^{\dag
},c_{l,\sigma }=\sqrt{\frac{\kappa }{\lambda }}a_{l,\sigma },  \notag \\
\bar{d}_{l,\sigma } &=&b_{l,\sigma }^{\dag },d_{l,\sigma }=b_{l,\sigma },
\label{trans}
\end{eqnarray}%
which obey the commutative relations%
\begin{eqnarray}
\left\{ c_{l,\sigma },\bar{c}_{l^{\prime },\sigma ^{\prime }}\right\}
&=&\delta _{ll^{\prime }}\delta _{\sigma \sigma ^{\prime }},\left\{ \bar{c}%
_{l,\sigma },\bar{c}_{l^{\prime },\sigma ^{\prime }}\right\} =\left\{
c_{l,\sigma },c_{l^{\prime },\sigma ^{\prime }}\right\} =0,  \notag \\
\left\{ d_{l,\sigma },\bar{d}_{l^{\prime },\sigma ^{\prime }}\right\}
&=&\delta _{ll^{\prime }}\delta _{\sigma \sigma ^{\prime }},\left\{ \bar{d}%
_{l,\sigma },\bar{d}_{l^{\prime },\sigma ^{\prime }}\right\} =\left\{
d_{l,\sigma },d_{l^{\prime },\sigma ^{\prime }}\right\} =0.  \notag \\
&&
\end{eqnarray}%
The transformation in Eq. (\ref{trans}) is essentially a similarity
transformation with singularities at $\lambda =0$ and $\kappa =0$, beyond
which it allows us to rewrite the Hamiltonian in the form%
\begin{equation}
H_{\mathrm{AB}}^{\prime }=\sum_{\sigma =\uparrow ,\downarrow }\sum_{l}\left[
\sqrt{\lambda \kappa }\bar{d}_{l,\sigma }c_{l,\sigma }+\sqrt{\lambda \kappa }%
\bar{c}_{l,\sigma }d_{l,\sigma }\right] .
\end{equation}%
So far, $H_{\mathrm{AB}}^{\prime }$ has become a Hermitian Hamiltonian,
which allows us to employ the perturbation method to get the effective
Hamiltonian
\begin{eqnarray}
\left( H_{\mathrm{AB}}^{\prime }\right) ^{\mathrm{eff}} &=&-\frac{8}{U}%
\sum_{l}(-\frac{\kappa ^{2}\eta _{\mathrm{A},l}^{-}\eta _{\mathrm{B},l}^{+}}{%
2} \\
&&-\frac{\lambda ^{2}\eta _{\mathrm{A},l}^{+}\eta _{\mathrm{B},l}^{-}}{2}%
+\lambda \kappa \eta _{\mathrm{A},l}^{z}\eta _{\mathrm{B},l}^{z}-\frac{%
\lambda \kappa }{4}).  \notag
\end{eqnarray}%
Although the above solution is only true for nonzero $\lambda $, one can
extrapolate the approximate solution at $\lambda =0$ by taking $\lambda
\rightarrow 0$. In the limit of zero $\lambda $, we have $H_{\mathrm{AB}%
}^{\prime }\rightarrow H_{\mathrm{AB}}$ and $\left( H_{\mathrm{AB}}^{\prime
}\right) ^{\mathrm{eff}}\rightarrow H_{\mathrm{AB}}^{\mathrm{eff}}.$\newline
\bigskip {\ \ }


\begin{thebibliography}{99}

\bibitem{Hubbard} J. Hubbard, Electron correlations in narrow energy bands,
Proc. R. Soc. A \textbf{276}, 238 (1963).

\bibitem{Jochim} {{S. Jochim, M. Bartenstein, A. Altmeyer, G. Hendl, S.
Riedl, C. Chin, }J. Hecker Denschlag, and R. Grimm, Bose-Einstein
Condensation }of Molecules,{\ Science \textbf{302}, 2101 (2003).}

\bibitem{Greiner} {M. Greiner, C. A. Regal, and D. S. Jin, Emergence of a
molecular }Bose--Einstein condensate from a Fermi gas, {Nature (London)
\textbf{426}, 537 (2003). }

\bibitem{Matthew} M. P. A. Fisher, P. B. Weichman, G. Grinstein, and D. S.
Fisher, Boson localization and the superfluid-insulator transition, Phys.
Rev. B \textbf{40}, 546 (1989).
\bibitem{Antoine} A. Georges, G. Kotliar, W. Krauth, and
M. J. Rozenberg, Dynamical mean-field theory of strongly correlated
fermion systems and the limit of infinite dimensions, Rev. Mod. Phys.
\textbf{68}, 13 (1996).
\bibitem{Keimer} B. Keimer, S. A. Kivelson, M. R. Norman, S. Uchida, and J.
Zaanen, From quantum matter to high-temperature superconductivity in copper
oxides, Nature {\ (London)} \textbf{518,} 179 (2015).
\bibitem{Patrick} P. A. Lee, N. Nagaosa, and X.-G. Wen, Doping a mott
insulator: Physics of high-temperature superconductivity, Rev. Mod. Phys.
\textbf{78,} 17 (2006).
\bibitem{CNY} C. N. Yang, $\eta $ pairing and off-diagonal long-range order
in a hubbard model, Phys. Rev. Lett. \textbf{63,} 2144 (1989).
\bibitem{Choi} S. Choi, J. Choi, R. Landig, G. Kucsko, H. Zhou, J. Isoya, F.
Jelezko, S. Onoda, H. Sumiya, V. Khemani, C. v. Keyserlingk, N. Y. Yao, E.
Demler, and M. D. Lukin, Observation of discrete time-crystalline order in a
disordered dipolar many-body system, Nature \textbf{543}, 221 (2017).
\bibitem{Else} D. V. Else, B. Bauer, and C. Nayak, Floquet time crystals,
Phys. Rev. Lett. \textbf{117}, 090402 (2016).
\bibitem{Khemani} V. Khemani, A. Lazarides, R. Moessner, and S. L. Sondhi,
Phase structure of driven quantum systems, Phys. Rev. Lett. \textbf{116},
250401 (2016).
\bibitem{Lindner} N. H. Lindner, G. Refael, and V. Galitski, Floquet
Topological Insulator in Semiconductor Quantum Wells, Nat. Phys. \textbf{7},
490 (2011).

\bibitem{Kaneko} T. Kaneko, T. Shirakawa, S. Sorella, and S. Yunoki,
Photoinduced $\eta $ Pairing in the Hubbard Model, Phys. Rev. Lett. \textbf{%
122}, 077002 (2019).

\bibitem{Tindall} J. Tindall, B. Bu\v{c}a, J.\thinspace R. Coulthard, and D.
Jaksch, Heating-Induced Long-Range \ $\eta $ Pairing in the Hubbard Model,
Phys. Rev. Lett. \textbf{123}, 030603 (2019).

\bibitem{ZXZPRB2} X. Z. Zhang and Z. Song, $\eta $-pairing ground states in
the non-Hermitian Hubbard model, Phys. Rev. B \textbf{103}, 235153 (2021).

\bibitem{YXMPRA} X. M. Yang and Z. Song, Resonant generation of a p-wave
Cooper pair in a non-Hermitian Kitaev chain at the exceptional point, Phys.
Rev. A \textbf{102}, 022219 (2020).%

\bibitem{QD1} M. Rigol, V. Dunjko, and M. Olshanii, Thermalization and Its
Mechanism for Generic Isolated Quantum Systems, Nature  (London) \textbf{452}, 854
(2008).

\bibitem{QD2} J. Eisert, M. Friesdorf, and C. Gogolin, Quantum Many-Body
Systems out of Equilibrium, Nat. Phys. \textbf{11}, 124 (2015).

\bibitem{QD3} M. Heyl, Dynamical quantum phase transitions: A brief survey,
Europhys. Lett. \textbf{125}, 26001 (2019).

\bibitem{YXMPRB} X. M. Yang and Z. Song, Quantum mold casting for
topological insulating and edge states, Phys. Rev. B \textbf{103}, 094307
(2021).

\bibitem{Berry2004} M. V. Berry, Physics of nonhermitian degeneracies,
Czech. J. Phys. \textbf{54}, 1039 (2004).

\bibitem{Heiss2012} W. D. Heiss, The physics of exceptional points, J. Phys.
A: Math. Theor. \textbf{45}, 444016 (2012).

\bibitem{Miri2019} M.-A. Miri and A. Al\'{u}, Exceptional points in optics
and photonics, Science \textbf{363}, eaar7709 (2019).

\bibitem{Zhang2020} X. Zhang and J. Gong, Non-Hermitian Floquet topological
phases: Exceptional points, coalescent edge modes, and the skin effect,
Phys. Rev. B \textbf{101}, 045415 (2020).

\bibitem{JL} L. Jin, P. Wang, and Z. Song, Su-Schrieffer-Heeger chain with
one pair of $\mathcal{PT}$-symmetric defects, Sci. Rep. \textbf{7,} 5903
(2017).

\bibitem{WPARXIV} P. Wang, K. L. Zhang, and Z. Song, Transition from
degeneracy to coalescence: theorem and applications, Phys. Rev. B \textbf{104%
}, 245406 (2021).

\bibitem{Kato} T. Kato, Perturbation theory of linear operator (Springer,
Berlin, 1966).

\bibitem{Muller} M. Muller and I. Rotter, Exceptional points in open quantum
systems, J. Phys. A: Math. Theor. \textbf{41}, 244018 (2008).

\bibitem{Moiseyev} N. Moiseyev, Non-Hermitian Quantum Mechanics (Cambridge
University Press, Cambridge, UK, 2011).

\bibitem{Emil} Emil J. Bergholtz, Jan Carl Budich, and Flore K. Kunst,
Exceptional topology of non-Hermitian systems, Rev. Mod. Phys. \textbf{93},
015005 (2021).

\bibitem{Doppler2016} J. Doppler, A. A. Mailybaev, J. B\"{o}hm, U. Kuhl, A.
Girschik, F. Libisch, T. J. Milburn, P. Rabl, N. Moiseyev, and S. Rotter,
Dynamically encircling an exceptional point for asymmetric mode switching,
Nature \textbf{537}, 76 (2016).

\bibitem{Xu2016} H. Xu, D. Mason, L. Jiang, and J. G. E. Harris, Topological
energy transfer in an optomechanical system with exceptional points, Nature
\textbf{537}, 80 (2016).

\bibitem{Assawaworrarit2017} S. Assawaworrarit, X. Yu, and S. Fan, Robust
wireless power transfer using a nonlinear parity--time-symmetric circuit,
Nature \textbf{546}, 387 (2017).

\bibitem{Wiersig2014} J. Wiersig, Enhancing the sensitivity of frequency and
energy splitting detection by using exceptional points: application to
microcavity sensors for single-particle detection, Phys. Rev. Lett. \textbf{%
112}, 203901 (2014).

\bibitem{Wiersig2016} J. Wiersig, Sensors operating at exceptional points:
general theory, Phys. Rev. A \textbf{93}, 033809 (2016).

\bibitem{Hodaei2017} H. Hodaei, A. U. Hassan, S. Wittek, H. Garcia-Gracia,
R. El-Ganainy, D. N. Christodoulides, and M. Khajavikhan, Enhanced
sensitivity at higher-order exceptional points, Nature \textbf{548}, 187
(2017).

\bibitem{Chen2017} W. Chen, \c{S}. Kaya \"{O}zdemir, G. Zhao, J. Wiersig,
and L. Yang, Exceptional points enhance sensing in an optical microcavity,
Nature \textbf{548}, 192 (2017).

\bibitem{Ding2016} K. Ding, G. Ma, M. Xiao, Z. Q. Zhang, and C. T. Chan,
Emergence, coalescence, and topological properties of multiple exceptional
points and their experimental realization, Phys. Rev. X \textbf{6}, 021007
(2016).

\bibitem{Xiao2019} Y.-X. Xiao, Z.-Q. Zhang, Z. H. Hang, and C. T. Chan,
Anisotropic exceptional points of arbitrary order, Phys. Rev. B \textbf{99},
241403(R) (2019).

\bibitem{Pan2019} L. Pan, S. Chen, and X. Cui, Interacting non-Hermitian
ultracold atoms in a harmonic trap: Two-body exact solution and a high-order
exceptional point, Phys. Rev. A \textbf{99}, 063616 (2019).
\end{thebibliography}
\end{document}